# On the entropy of the Viana-Bray model

**J.R.L. de Almeida**
**Departamento de Física, Universidade Federal de Pernambuco**
**50670-901, Recife-PE, Brazil**

**ABSTRACT**

The entropy of the Viana-Bray model at zero temperature and external field is calculated within the solution which takes into account only delta functions for the global order parameter P(h). It is shown that such solution is unsatisfactory both from the viewpoint of stability analysis and for not reproducing the well known Sherrington-Kirkpatrick result in the large connectivity limit thus pointing out the relevance of considering solutions with continuous part in P(h) for such model and possibly related models.

Recently a lot of work has been devoted to the study of finite connectivity spin glasses and related models associated to the satisfiability problem ( see [1] and references therein ). A prototype of these finite connectivity models is the Viana-Bray (VB) model [2] designed to study diluted magnetic systems and applicable to optimization problems [3], specially the 2-SAT problem [4]. One of the most usual analytical technique used to treat these problems is the replica method which for the Sherrington-Kirkpatrick (SK) model [5] yields a negative entropy at low temperatures within the replica symmetric (RS) solution , signaling a problem for the unbroken symmetry ansatz. Eventually, stability study of the RS solution [6] pointed out the necessity to introduce broken replica simmetry solution which led to a physically acceptable solution [7]. Although the VB model has been intensely studied, the stability study of its solutions has been thoroughly analyzed mainly close to the percolation threshold both for the global order parameter with only discrete components [8] or including continuous component [9]. In this note the entropy of the VB model for the discrete component solution at zero temperature and external field is considered and shown to be rather unsatisfactory for it is plainly unstable against longitudinal variations and moreover does not yield SK's result in the large coordination limit. The VB model is described by the Hamiltonian

$$H = \sum_{(ij)} J_{ij}\, \sigma_i \sigma_j \qquad (1)$$

where $\sigma_i = \pm 1$ $(i = 1, 2, ...., N)$, and the $J_{ij}$'s are infinite-ranged random interactions with probability distribution given by

$$P(J_{ij}) = (1 - p/N)\delta(J_{ij}) + (p/N)f(J_{ij}). \tag{2}$$

It thus may describe a highly diluted system with average connectivity $p$. The distribution of the active bonds shall here be taken as a bimodal: $f(J_{ij}) = (\delta(J_{ij} - 1) + \delta(J_{ij} + 1))/2$. The variational free energy $f$ within the replica method takes the form [ 8,9]

$$-\beta fn = -p/2 - p Tr_\sigma \{g_n(\sigma_\alpha) \exp[g_n(\sigma_\alpha)]\}/2 Tr_\sigma \exp[g_n(\sigma_\alpha)] +$$
$$+ \ln Tr_\sigma \exp[g_n(\sigma_\alpha)] \tag{3}$$

where $g_n(\sigma_\alpha)$ is a generalized global order parameter, with $2^n$ components (n is the number of replicas which is made go to zero at the end of the calculation) and $\beta = 1/T$ the inverse temperature. It is related to the functional probability distribution of the local fields by a Fourier-like transformation [10] . For our purposes it is more convenient express $f$ in terms of the probability distribution of the local fields $P(h)$ as in [3], valid for replica symmetric solution, which reads

$$f = -\frac{p}{2\beta} \int dJ f(J) \ln[\cosh(\beta J)] - \frac{1}{\beta} \int dh P(h) \ln[2 \cosh(\beta h)] +$$
$$+ \frac{p}{2\beta} \int dJ dh dh' f(J) P(h) P(h') \ln[1 + \tanh(\beta J) \tanh(\beta h) \tanh(\beta h') -$$

4$$-\frac{p}{2\beta}\int dJ dh f(J)P(h)\ln[1-\tanh^2(\beta J)\tanh^2(\beta h)] \qquad (4)$$

The global order parameters $g_n(\sigma_\alpha)$ or $P(h)$ obey their equation of motion [3,8,9] which for the latter is

$$P(h) = \int \frac{dy}{2\pi}\exp\left\{-iyh - p + p\int dJ f(J)\int dx P(x)\exp\left[\frac{iy}{\beta}\tanh[\tanh(\beta J)\tanh(\beta x)]\right]\right\} \qquad (5)$$

and the relationship between the two order parameters is

$$P(h) = \int dy \exp[-iyh + g(y)] \qquad (6)$$

Depending on the quantity of interest it may be easier to work with one or the other of these two parameters although the field distribution has a seemingly clear physical appeal. To infer the form of the general replica symmetric solution is seems better to work with g(y) as originally put forward by Katsura [11] for the Bethe lattice and extended for the VB model by de Almeida et al [9]. The simplest solution at very low temperatures is the one assuming that [3]

$$P(h) = (1-Q)\delta(h) + \sum_{l=1}^{\infty} p_l^+ \delta(h-l) + \sum_{l=1}^{\infty} p_l^- \delta(h+l) \qquad (7)$$

which we shall take as the one valid close to zero temperature, except for vanishingly small exponential corrections, for as shown in [12] it is found that $P(h)$ below the spin glass temperature is almost constant. Using the above equation of motion for $P(h)$ it is straigthforward to show that the parameters in (7) satisfy

$$1 - Q = \exp(-pQ)I_0(pQ) \qquad (8)$$

and

$$p_l^\pm = \exp(-pQ)I_l(pQ) \tag{9}$$

where $I_l(x)$ are modified Bessel functions of order $l$. From equation (4) it is easy to obtain the entropy per spin $s = -df/dT$, at $T = 0$, which reads

$$s = -\frac{p}{2}\ln(2)(1-Q)^2 + (1-Q)\ln(2) + \frac{p}{2}\ln(2)(p_1^+ + p_1^-)(1-Q) - \frac{p}{4}(p_1^+ + p_1^-)^2 \ln(3/4)$$

(10)

and as $I_l(x) \cong \exp(x)/\sqrt{2\pi x}$, for large $x$, the s limiting value as $p \to \infty$ is positive and given by $s \cong 0.117$. It should not be surprising that the entropy is positive in view of the discrete nature of the solution (7) but in this limit SK's result should be recovered which is obviously not the case. Another way to study the validity of the solution (7) is to consider the fluctuations of the variational free energy around it as carried out in [8] and [9]. This has been done thoroughly for $p$ close to the percolation threshold [8] and the results show that the solution is unstable even for just longitudinal variations. However, one may easily show that (7) is an unstable solution for any $p$, at zero temperature, by considering the $\lambda$–eigenvalue equations in [8,9] which for longitudinal variations in $g_n(\sigma_\alpha)$ is

$$f(x) = -[p + g(x)]f(0) + p\int dy K(x,y)(\lambda + e^{g(y)})f(y) \tag{11}$$

where

$$K(x,y) = \int dJ f(J) \int \frac{du}{2\pi} \exp\left[-iuy + \frac{ix}{\beta}\tanh[\tanh(\beta J)\tanh(\beta u)]\right] \tag{12}$$

and for a stable solution, at least for longitudinal variations, $\lambda > 0$ for all possible solutions of (11). For the solution (7), we find the eigenvalues $\lambda_1 = 1/p, \lambda_2 = (1-p)/p(1-Q)$, and $\lambda_3 = 2(1-p)/p$ clearly showing that the solution

(7) is unstable for $p>1$. So it seems that the continuous part of $P(h)$ must be taken in account for the VB model if well known results are to be recovered and possibly similar models specially those relevant to the K-SAT problem. **Acknowledgment:** the author is grateful to PRONEX/MCT for partial financial support.